\documentclass{moriond}
\usepackage{amsmath}

\def\be{\begin{equation}}
\def\ee{\end{equation}}
\def\bea{\begin{eqnarray}}
\def\eea{\end{eqnarray}}

\newcommand{\Photo}{}

\begin{document}
\vspace*{4cm}
\title{Dark Matter searches at CMS and ATLAS}

\author{ Danyer P{\' e}rez Ad{\' a}n, on behalf of the ATLAS and CMS Collaborations }

\address{Deutsches Elektronen-Synchrotron (DESY), CMS group,\\
Notkestra{\ss}e 85, 22607 Hamburg, Germany}

\maketitle\abstracts{
A glimpse into the most recent searches for Dark Matter at the LHC performed by the ATLAS and CMS collaborations is presented. The results covered in this document are all based on Run-2 proton-proton collision data recorded at $\sqrt{s}=13\text{ TeV}$ and corresponding to an integrated luminosity of just under 140 fb\textsuperscript{-1}.
}

\footnote[0]{Copyright [2022] CERN for the benefit of the [ATLAS and CMS Collaborations]. Reproduction of this article or parts of it is allowed as specified in the CC-BY-4.0 license.}

\vspace{-0.8cm}

\section{Introduction}
\label{sec:intro}

Various astrophysical observations~\cite{Bertone:2004pz} point to the existence of large amounts of Dark Matter (DM) in the universe. There are many models suggesting that the nature of the DM could be particle-like, situating it in the context of new physics beyond the Standard Model (SM)~\cite{Abercrombie:2015wmb}. The number of possible candidates embraced by these theories is significant, however, the `dark' particles have a few distinctive properties that are possible to exploit in the hunt for DM. 

At colliders, such as the CERN LHC, the DM particles are expected to be produced after the collisions via the mediated interaction with SM particles. The main characteristic of the DM particles produced in the final state is that they are effectively invisible to experiments located at the LHC. The invisible particles would normally traverse the CMS~\cite{CMS:2008xjf} and ATLAS~\cite{ATLAS:2008xda} detectors completely undetected and leaving behind an unbalanced momentum in the transverse plane of the proton-proton collision. The above makes this quantity, known as missing transverse momentum ($\vec{p}^{\text{ miss}}_{T}$), the main probed observable when searching for DM production at the LHC.

\begin{figure}[!ht]
\centering
\includegraphics[width=0.48\textwidth]{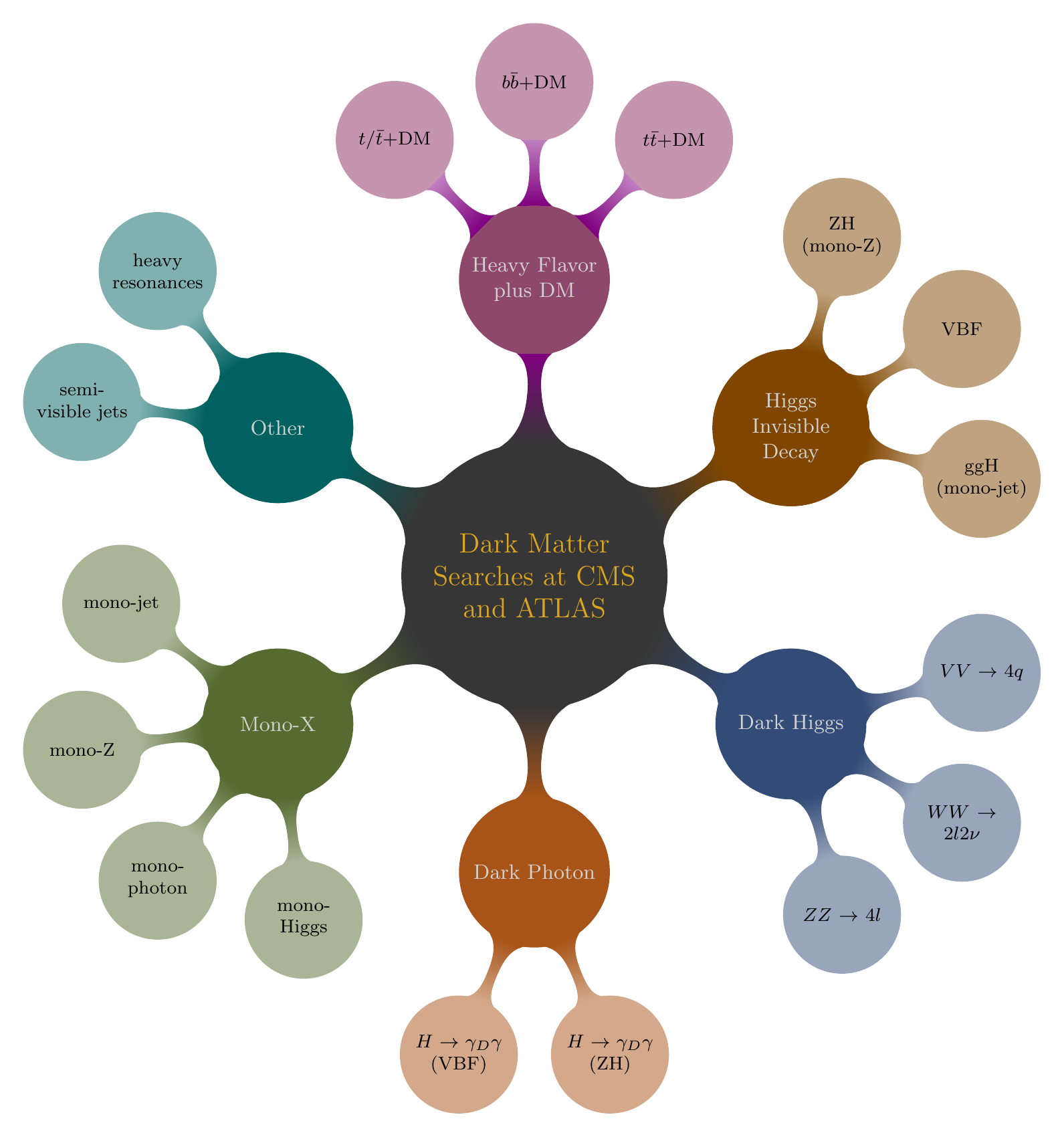}
\vspace{-0.3cm}
\caption[]{ A sketch of the most common type of DM searches performed at ATLAS and CMS divided into a few subgroups, depending on the particles recoiling against (accompanying) the invisible particles.}
\label{fig:search_program}
\end{figure}

Any remaining product of the collision becomes useful in the identification of events with the presence of DM particles, hence the reason why many of these searches are commonly classified as $\vec{p}^{\text{ miss}}_{T}+X$ analyses, where $X$ represents any possible SM particle accompanying the DM candidate. The DM search program at ATLAS and CMS can be represented in a sketch such as the one in Fig.~\ref{fig:search_program}, which is not intended to be exhaustive but accommodates the most standard searches in a few categories. The heavy flavor category is motivated by models in which the coupling between the mediators and the SM fermions is of Yukawa type, thus favoring the associated production of DM with top-quarks or b-quarks. The Higgs to invisible decay group focuses on theories that incorporate the new hidden sector through interactions with the scalar field, which could potentially produce a sizable decay rate to DM particles of the newly discovered boson. The dark Higgs and dark photon searches are more specific to these two widely explored models and are usually optimized for specific production and decay channels. Another big sector is the one constituted by the mono-X analyses, which are mainly looking for one single SM particle recoiling against DM; those tend to be less model-dependent searches, as some of the X particles (e.g. jets) can be copiously produced from the initial state radiation in proton-proton collisions. Other more exotic kinds of DM searches are grouped in the last remaining category, which comprises a large number of theoretical models.

\section{Search for DM with an energetic jet and missing transverse momentum}\label{sec:mono_jet}

\begin{figure}[!ht]
\centering
\raisebox{0.3\height}{\includegraphics[width=0.25\textwidth]{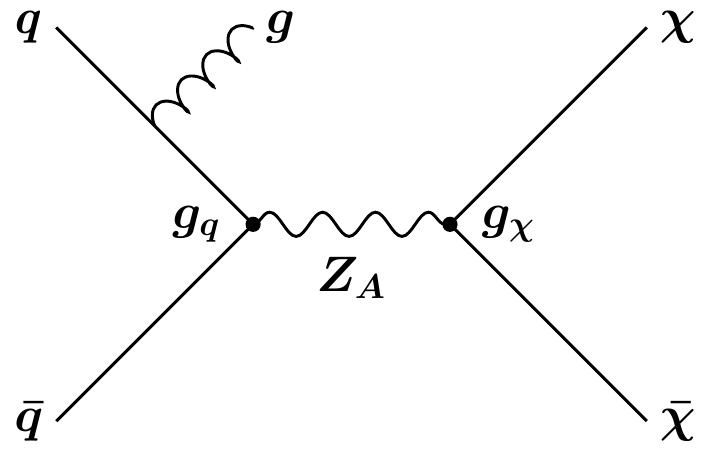}}
\raisebox{0.3\height}{\includegraphics[width=0.23\textwidth]{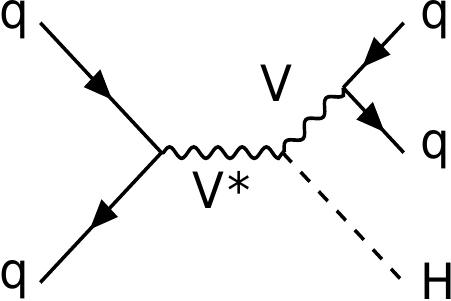}}
\hspace{1cm}
\includegraphics[width=0.35\textwidth]{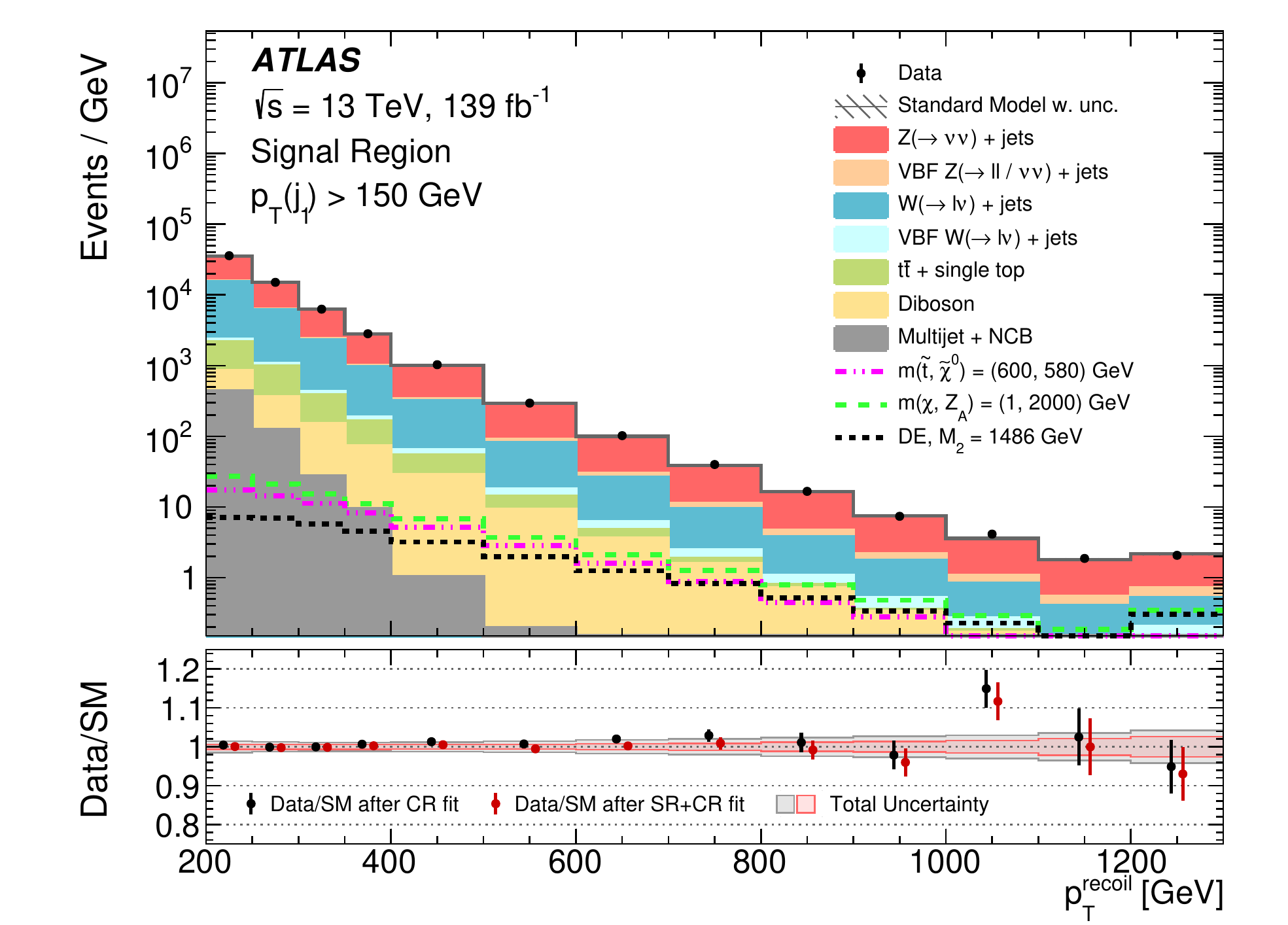}
\vspace{-0.3cm}
\caption[]{ Feynman diagrams representing two of the processes that could lead to mono-jet signatures: initial state emission of a jet in a simplified DM model with a axial-vector mediator exchanged in the s-channel (left)~\cite{ATLAS:2021kxv} and Higgs Strahlung production for hadronic decay of the vector boson and consequently invisible decay of the Higgs boson (center)~\cite{CMS:2021far}. Distribution of the final discriminant observable for the ATLAS analysis (right)~\cite{ATLAS:2021kxv}.}
\label{fig:monojet:discriminant}
\end{figure}

One of the analyses belonging to the mono-X group is the mono-jet search, which looks for an energetic hadronic jet recoiling against a large amount of $\vec{p}^{\text{ miss}}_{T}$. Fig.~\ref{fig:monojet:discriminant} (left and center) shows two representative examples of processes that could exhibit such experimental signature; one as a consequence of an initial state emission of a jet in the context of a simplified DM model and the other as a result of an invisible decay of the Higgs boson in VH production.

The two searches performed by the CMS~\cite{CMS:2021far} and ATLAS~\cite{ATLAS:2021kxv} collaborations present, in general, a very similar analysis strategy. The main physics objects exploited in the selection are the jets, which are reconstructed using a distance parameter of $R=0.4$. In addition to this type of jets, the CMS analysis makes use of the so-called \textit{fat} jets, reconstructed with a larger cone of $R=0.8$. It also employs a deep neural network to differentiate between fat jets originating from a vector boson decay and those coming from QCD radiation; this is additionally used for event categorization in two regions with low and high purity of mono-V events. Events containing any $e$, $\mu$, $\tau$, or $\gamma$ are vetoed. A lower threshold is imposed on the variable $|\Delta\phi(\vec{p}^{\text{ jet}}_{T},\vec{p}^{\text{ miss}}_{T})|$, which is intended to reject QCD events with misreconstructed jets that originate a fictitious $\vec{p}^{\text{ miss}}_{T}$ along the jet momentum.

The main SM background processes contributing to the above selection are $Z(\rightarrow \bar{\nu}\nu)+\text{Jets}$ and $W(\rightarrow l\bar{\nu_{l}})+\text{Jets}$. The first process features an almost identical signature compared to the signal if a hard jet emission is produced in the event. Similarly, for the $W(\rightarrow l\bar{\nu_{l}})+\text{Jets}$ process, if the lepton flies outside the detector acceptance range, or if this particle is not reconstructed in the event, an apparent large $\vec{p}^{\text{ miss}}_{T}$ is perceived. These two backgrounds are estimated via subsidiary measurements in dedicated control regions (CRs) with two and one leptons, respectively. The devised CRs allow constraining the normalization of the above two processes, thus making possible the reduction of systematic uncertainties in the SRs.

A binned maximum likelihood fit is performed simultaneously across all signal regions (SRs) and CRs on the $\vec{p}^{\text{ miss}}_{T}$ distribution. Fig.~\ref{fig:monojet:discriminant} (right) illustrates the result obtained, evidencing that the search does not find any deviation from the SM expectation. Both ATLAS and CMS interpret this result in terms of the exclusion limits on the $\mathcal{B}(H\rightarrow\text{Inv})$, as well as on the parameters of simplified models with axial and pseudoscalar mediators. For $H\rightarrow\text{Inv}$, the observed (expected) limits obtained by CMS are $\mathcal{B}(H\rightarrow\text{Inv}) < 0.28 (0.25)$, whereas ATLAS reports $\mathcal{B}(H\rightarrow\text{Inv}) < 0.34 (0.39)$.

\section{Search for invisible decays of the Higgs boson in vector boson fusion}\label{sec:htoinv}

Undoubtedly, when it comes to $H \rightarrow \text{Inv}$ searches, the VBF production mode offers the greatest single-channel sensitivity due to its excellent balance between cross section and experimental signature. Recently, CMS has delivered the latest result on this channel~\cite{CMS:2022qva}, which is discussed in this section, but also ATLAS reported a similar search not long ago~\cite{ATLAS:2022yvh}.

The CMS analysis is divided into two categories, one with moderate $\vec{p}^{\text{ miss}}_{T}$ and another one with high $\vec{p}^{\text{ miss}}_{T}$. The jet selection focuses on having a pair of these falling in opposite hemispheres of the detector by requiring $\eta_{j_{1}} \cdot \eta_{j_{2}} < 0$. A large separation ($|\eta_{jj}|>1$) between the two jets is also imposed, as well as a substantial $|\Delta\phi(\vec{p}^{\text{ jet}}_{T},\vec{p}^{\text{ miss}}_{T})|$. A veto on all types of charged leptons and photons is applied.

The SM background is dominated by $Z(\rightarrow \bar{\nu}\nu)+\text{Jets}$ and $W(\rightarrow l\bar{\nu_{l}})+\text{Jets}$ processes with similar explanation as for the mono-jet (Sec.~\ref{sec:mono_jet}) case. To estimate these, CRs are defined requiring one or two charged leptons, but keeping the $\vec{p}^{\text{ miss}}_{T}$ and jet selection identical to the SRs.

A simultaneous binned maximum likelihood fit is performed on the $m_{jj}$ distribution, after which it was verified that the background describes relatively well the data. There is a small excess in some bins of the $m_{jj}$ distribution, but the significance of this disagreement is still low. The results are therefore interpreted in the context of $H \rightarrow \text{Inv}$ decays setting limits on the $\mathcal{B}(H\rightarrow\text{Inv})$ assuming SM Higgs cross sections. This yields an observed (expected) upper limit of $\mathcal{B}(H\rightarrow\text{Inv}) < 0.18 (0.10)$. The ATLAS analysis obtained an identical result for the expected upper limit but a slightly lower observed limit, setting $\mathcal{B}(H\rightarrow\text{Inv}) < 0.15 (0.10)$.

\section{Search for DM produced in association with a top quark and a W boson}\label{sec:tWDM}

\begin{figure}[!ht]
\centering
\raisebox{0.2\height}{\includegraphics[width=0.22\textwidth]{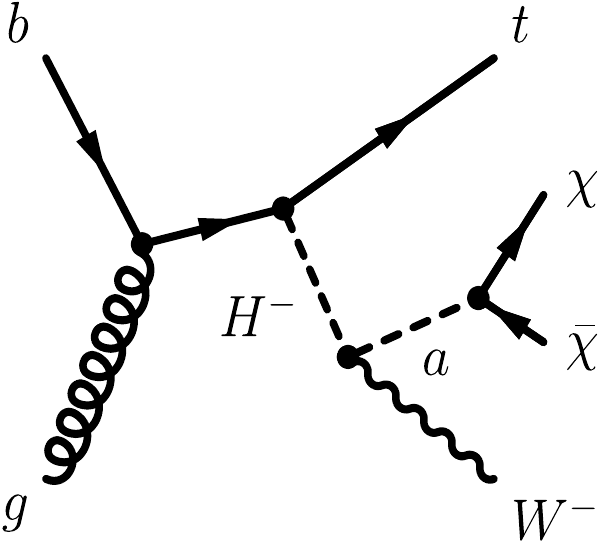}}
\includegraphics[width=0.36\textwidth]{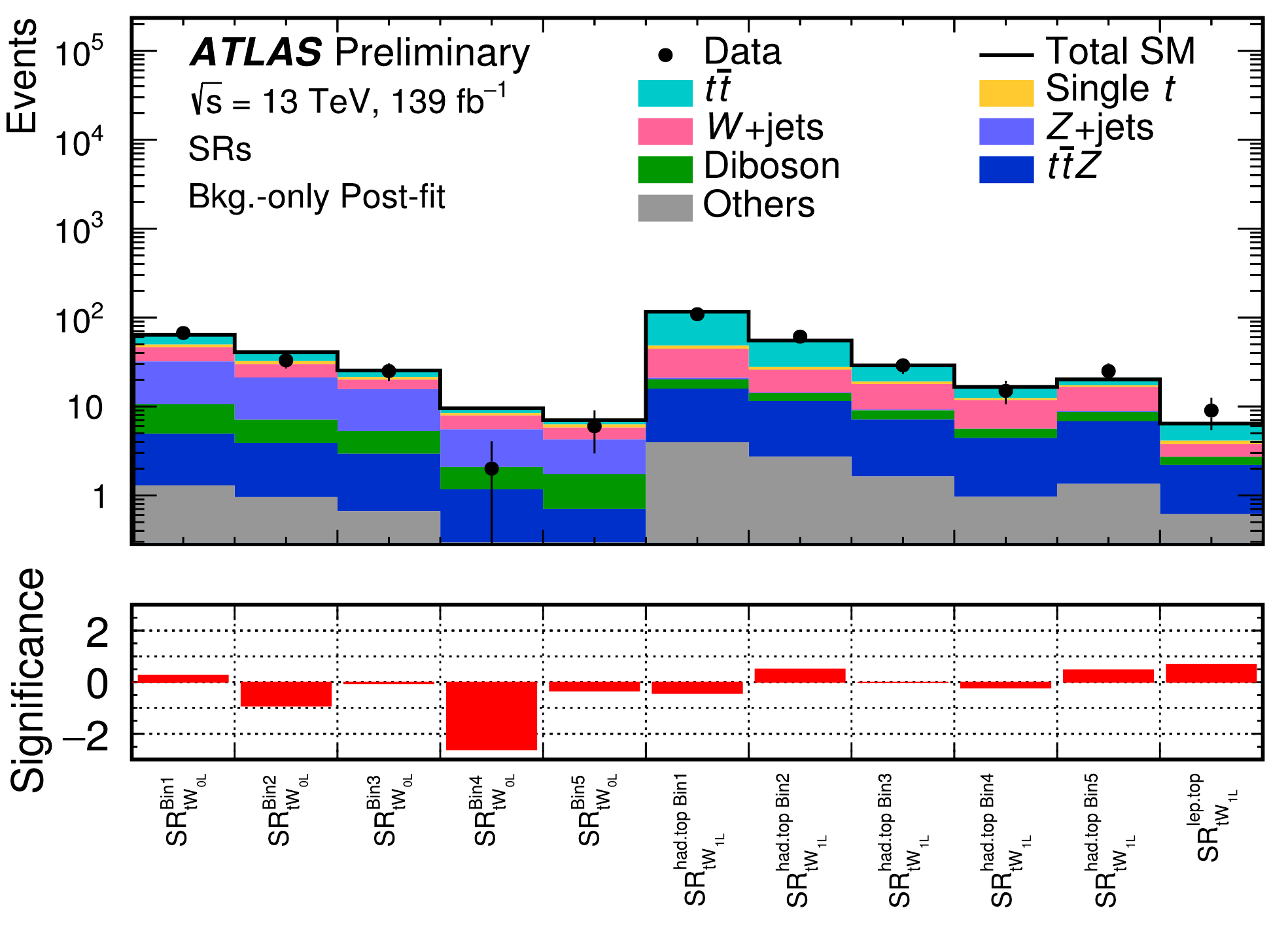}
\includegraphics[width=0.36\textwidth]{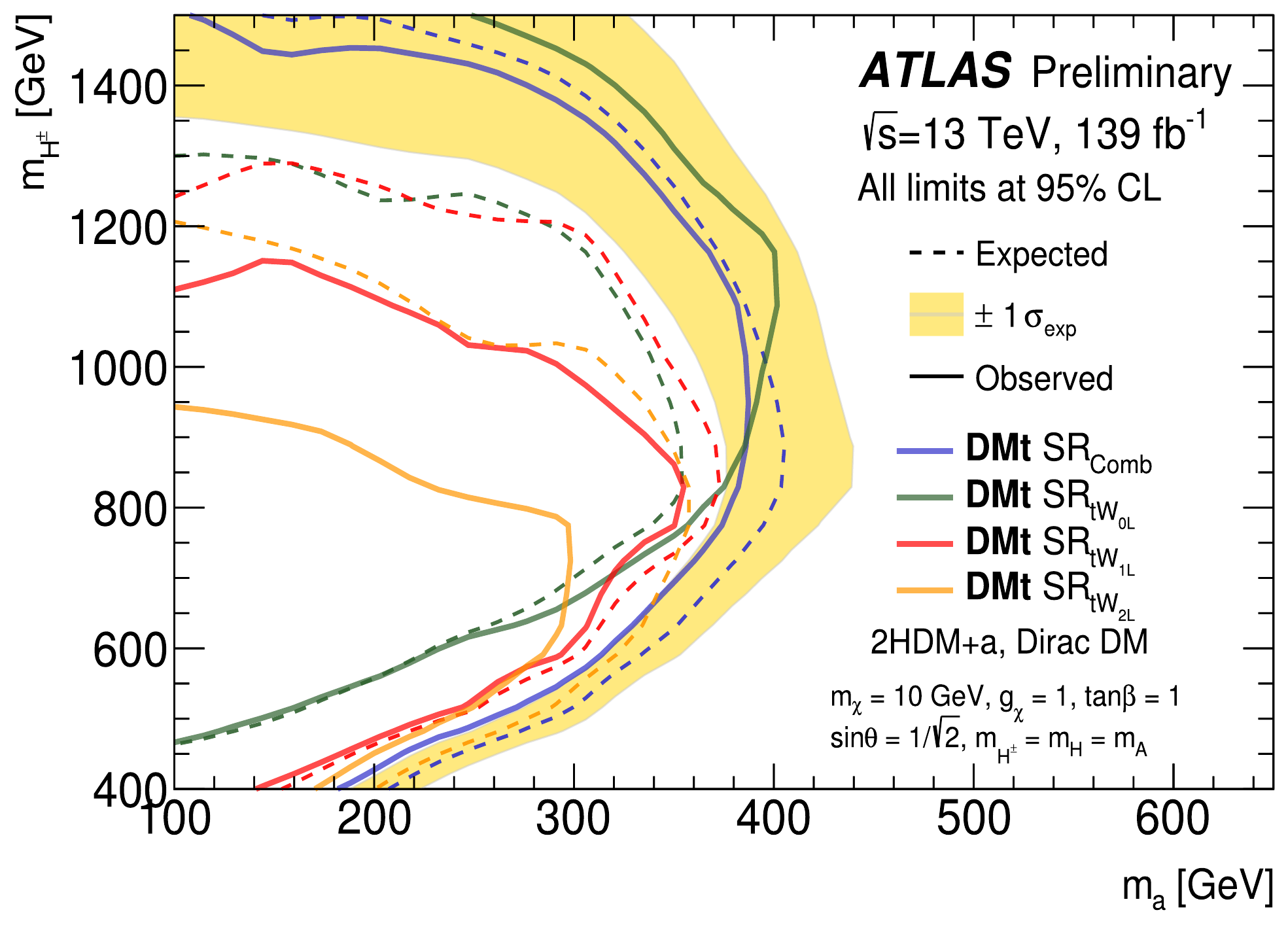}
\vspace{-0.3cm}
\caption[]{ Representative Feynman diagram in the 2HDMa model of the processes contributing to the final state here considered (left)~\cite{ATLAS-CONF-2022-012}. Final discriminant distribution after having performed a background-only fit to the data (center)~\cite{ATLAS-CONF-2022-012}. Exclusion limits in the 2D plane formed by the masses of $H^{\pm}$ and $a$ showing the results of the combination among all channels and their individual impact on the constraints (right)~\cite{ATLAS-CONF-2022-012}.}
\label{fig:tWDM:result}
\end{figure}

Looking for DM produced in association with heavy quarks is a regular undertaking when considering simplified models. However, in more complex scenarios such as the 2HDMa~\cite{LHCDarkMatterWorkingGroup:2018ufk} model, there exist more rich dynamics that can lead to changes in the search strategy usually adopted. That is the case of this search~\cite{ATLAS-CONF-2022-012}, performed by the ATLAS collaboration. If the charged Higgs states ($H^{\pm}$) are sufficiently heavy and the lightest pseudoscalar ($a$) is relatively light, there could be decays of the form $H^{\pm} \rightarrow W^{\pm}a $, where the $W$ and $a$ bosons would be boosted (Fig.~\ref{fig:tWDM:result} (left)). For hadronic decays of the $W$ boson, this poses a huge impact at the object reconstruction level.

The selection targets two main categories, the first one for events with zero leptons ($tW_{0l}$), and the second one for events with exactly one lepton ($tW_{1l}$). For both categories, the event selection requires a high $\vec{p}^{\text{ miss}}_{T}$, no presence of $\tau$ or $\gamma$ in the event, and at least one b-tagged jet. The main feature is the use of machine learning techniques to identify a fat jet consistent with the $W$ boson; in the $tW_{0l}$ category one of such objects is selected. In the $tW_{1l}$ category, the selection is further split depending on whether the lepton is coming from the $W$ boson or the top quark. One W-tagged fat jet is selected if the lepton comes from the top quark, otherwise, no W-tagged fat jet is selected.

The SM processes that prevail with this selection are the $Z+\text{Jets}$, $W+\text{Jets}$, $t\bar{t}$, and $t\bar{t}Z$. For each of these main backgrounds, a CR with the aim of constraining the normalization is constructed, using additional leptons in the event or cuts on some kinematic variables. 

The maximum likelihood fit is performed using several bins of the $\vec{p}^{\text{ miss}}_{T}$ distribution, varying the number of them depending on the category. The result of the background-only fit is shown in Fig.~\ref{fig:tWDM:result} (center), where one can see that there is no significant deviation in the data with respect to the SM prediction. The analysis here described was then combined with a previous ATLAS search~\cite{ATLAS:2020yzc} in the final state containing two leptons. The result of this combination is depicted in Fig.~\ref{fig:tWDM:result} (right), where a scan is performed in the 2D plane formed by $m_{a}$ and $m_{H^{\pm}}$.

\section{Search for DM produced in association with a dark Higgs boson}\label{sec:darkH}

\begin{figure}[!ht]
\centering
\raisebox{0.15\height}{\includegraphics[width=0.22\textwidth]{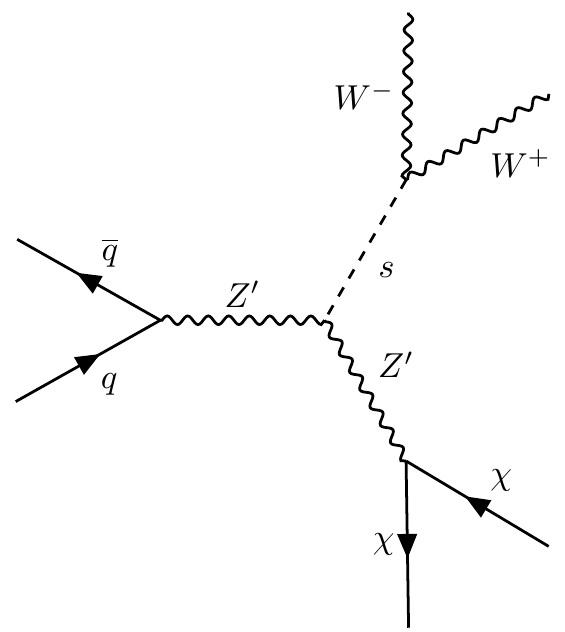}}
\hfill
\raisebox{0.0\height}{\includegraphics[width=0.31\textwidth]{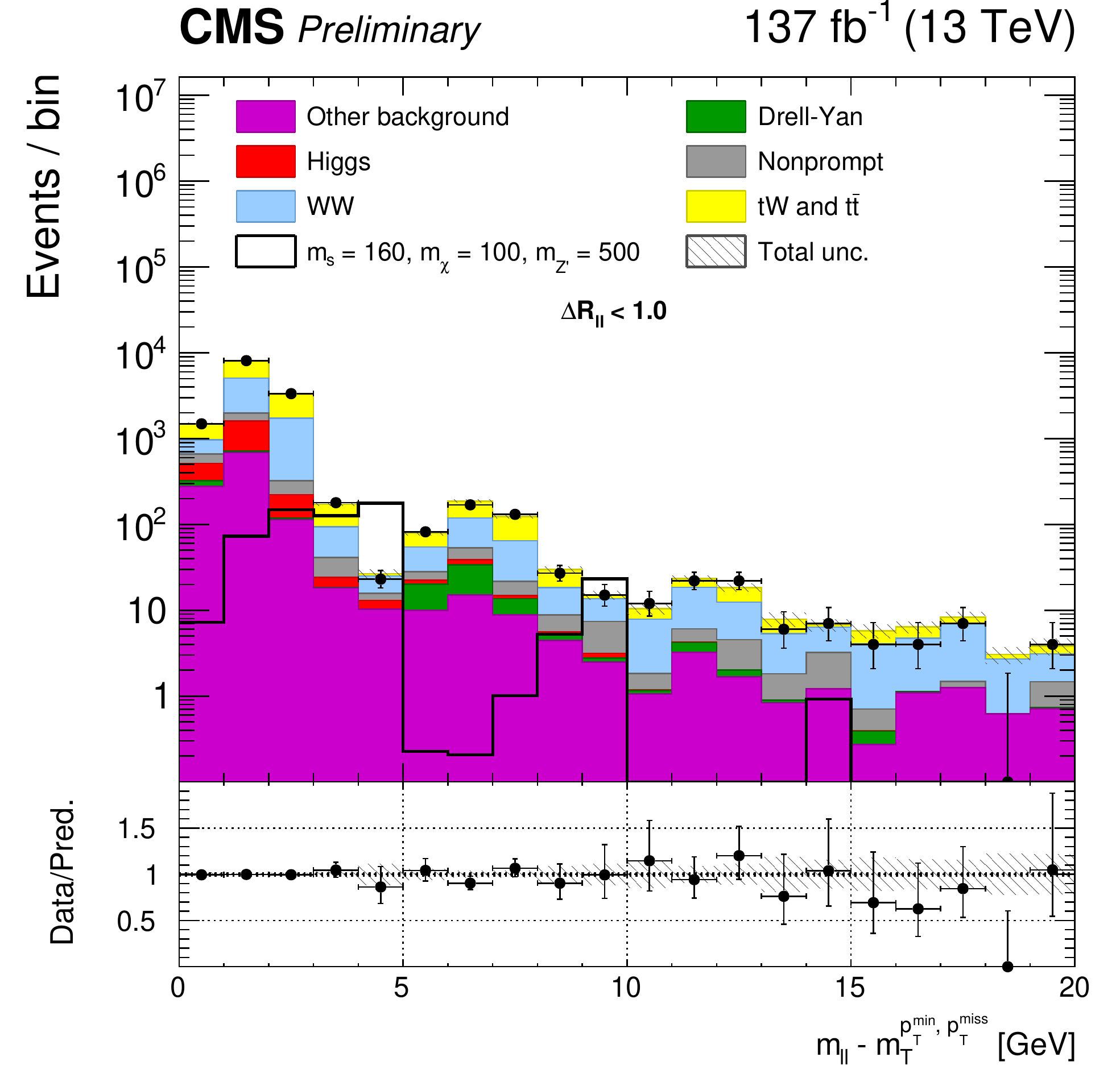}}
\hfill
\raisebox{0.0\height}{\includegraphics[width=0.33\textwidth]{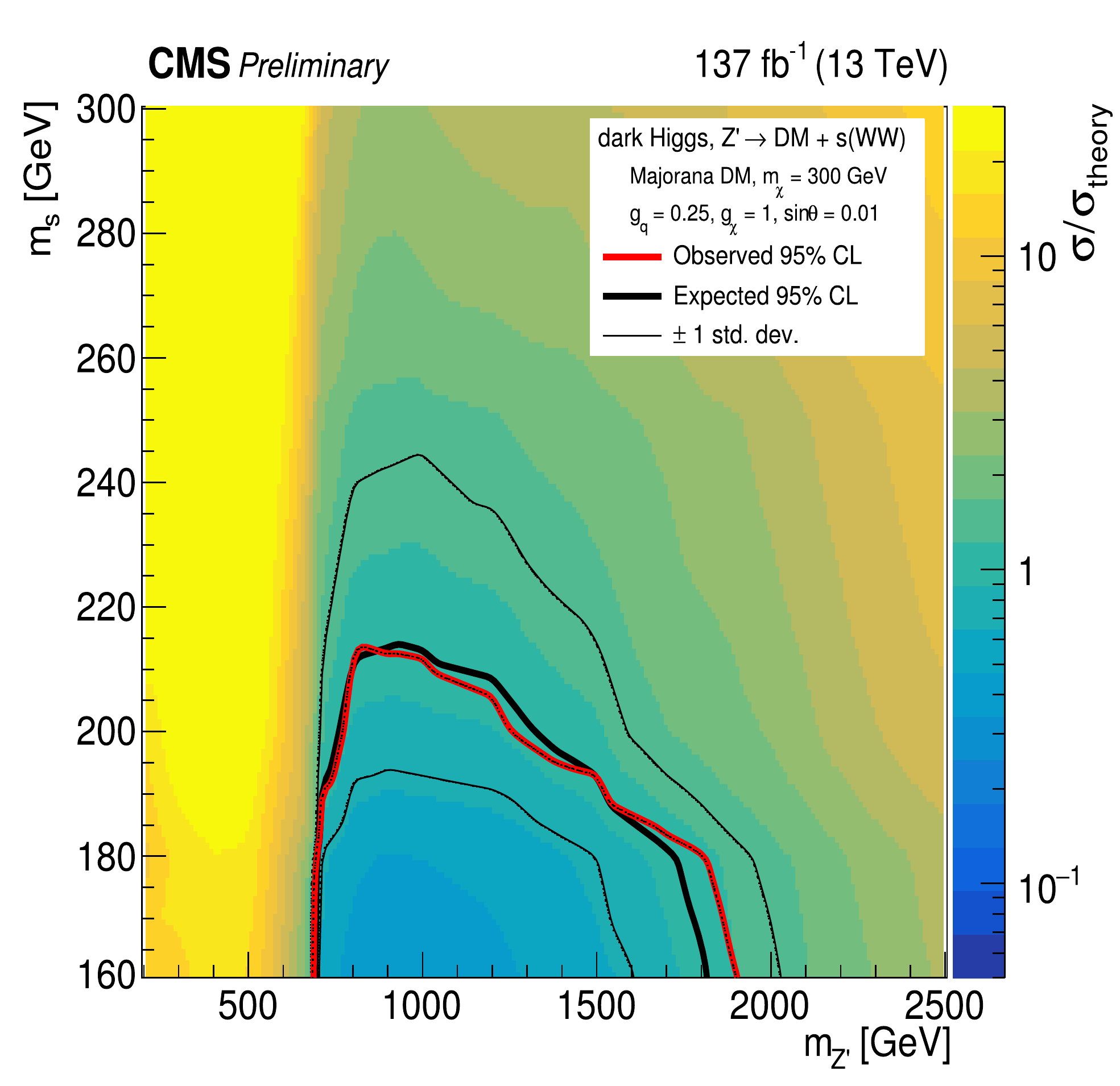}}
\vspace{-0.3cm}
\caption[]{ Representative Feynman diagram of the production of a dark Higgs (left)~\cite{CMS-PAS-EXO-20-013}. Final 1D (unrolled 2D) distribution on which the final fit is performed for one of the categories (center)~\cite{CMS-PAS-EXO-20-013}. Exclusion limits in the 2D plane formed by $m_{Z'}$ vs $m_{s}$ fixing the remaining model parameters (right)~\cite{CMS-PAS-EXO-20-013}.}
\label{fig:darkH:result}
\end{figure}

As opposed to most simplified DM models, the dark Higgs model offers an explanation of how the mass of the particles in the dark sector is acquired. The most recent search for DM targeting this specific model was delivered by the CMS collaboration~\cite{CMS-PAS-EXO-20-013}, looking for decays of the dark Higgs into $W^{+}W^{-}$ in the two lepton final state. Previously, ATLAS~\cite{ATLAS:2020fgc} had also performed a search in the fully hadronic final state, including both $W^{+}W^{-}$ and $ZZ$ intermediate resonances. The breaking of the additional $U(1)'$ gauge symmetry produces the coupling between the dark Higgs ($s$) and the incorporated gauge $Z'$ boson, thus leading to the signature represented in the Feynman diagram in Fig.~\ref{fig:darkH:result} (left). For masses of $s$ above the $W^{+}W^{-}$ threshold, this channel becomes the dominant decay mode, and it is, therefore, the focus of the present CMS analysis.

In this analysis, two opposite-charge leptons with different flavors are selected. To reduce the contribution from top quark processes, events containing any number of b-tagged jets are vetoed. The events are categorized based on the angular distance ($\Delta R_{ll}$) between the two leptons, resulting in three categories targeting different boosting regimes. A key observable in this search is the transverse mass of the trailing lepton and $\vec{p}^{\text{ miss}}_{T}$ system ($m^{l_{min},p^{\text{miss}}_{T}}_{T}$), which is used in combination with $m_{ll}$ to build the final distribution on which the fit is performed.

The modeling of the main backgrounds ($W^{+}W^{-}$, $t\bar{t}$, and $tW$) is mostly based on simulation, with additional regions to control their normalization. The rate for $W^{+}W^{-}$ is derived in a CR by requiring large $\Delta R_{ll}$, whereas for $t\bar{t}$ and $tW$ the selection allows for b-tagged jets.

The fit is performed on the binned 2D distribution formed by the variables $m^{l_{min},p^{\text{miss}}_{T}}_{T}$ and $m_{ll}$. The results are illustrated in Fig.~\ref{fig:darkH:result} (center), where the 2D template has been unrolled into a 1D distribution. There was no evidence of a discrepancy between the SM prediction and the data, and limits were set on the model parameters. Fig.~\ref{fig:darkH:result} (right) shows the excluded phase space for a specific configuration in which the mass of $s$ and $Z$ are varied simultaneously.

\section{Search for strongly coupled DM}\label{sec:svj}

One typical selection cut in the above searches using jets that is intended to remove the enormous QCD background is the requirement of these having a large $|\Delta\phi(\vec{p}^{\text{ jet}}_{T},\vec{p}^{\text{ miss}}_{T})|$. However, there are exotic models supposing the existence of hidden sectors with strong dynamics that can produce jets with both visible and invisible particles inside, known as semi-visible jets. In this case, the fraction of $\vec{p}^{\text{ miss}}_{T}$ coming from one of these jets will be relatively aligned with the direction of the visible jet, making the above mentioned selection cut very inefficient. That is the reason why the CMS collaboration recently embarked on a search~\cite{CMS:2021dzg} for this type of signature.

\begin{figure}[!ht]
\centering
\raisebox{0.0\height}{\includegraphics[width=0.33\textwidth]{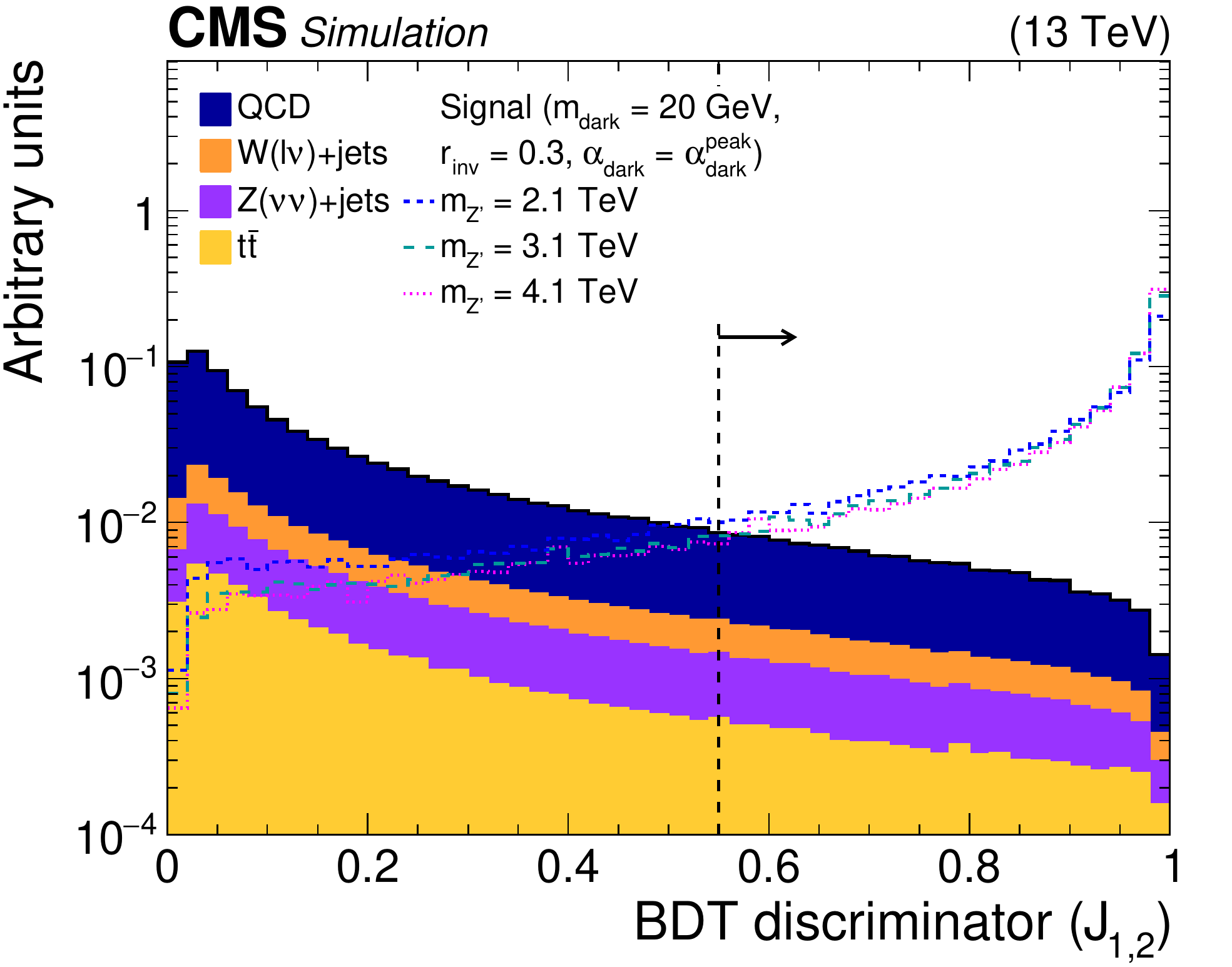}}
\raisebox{0.0\height}{\includegraphics[width=0.26\textwidth]{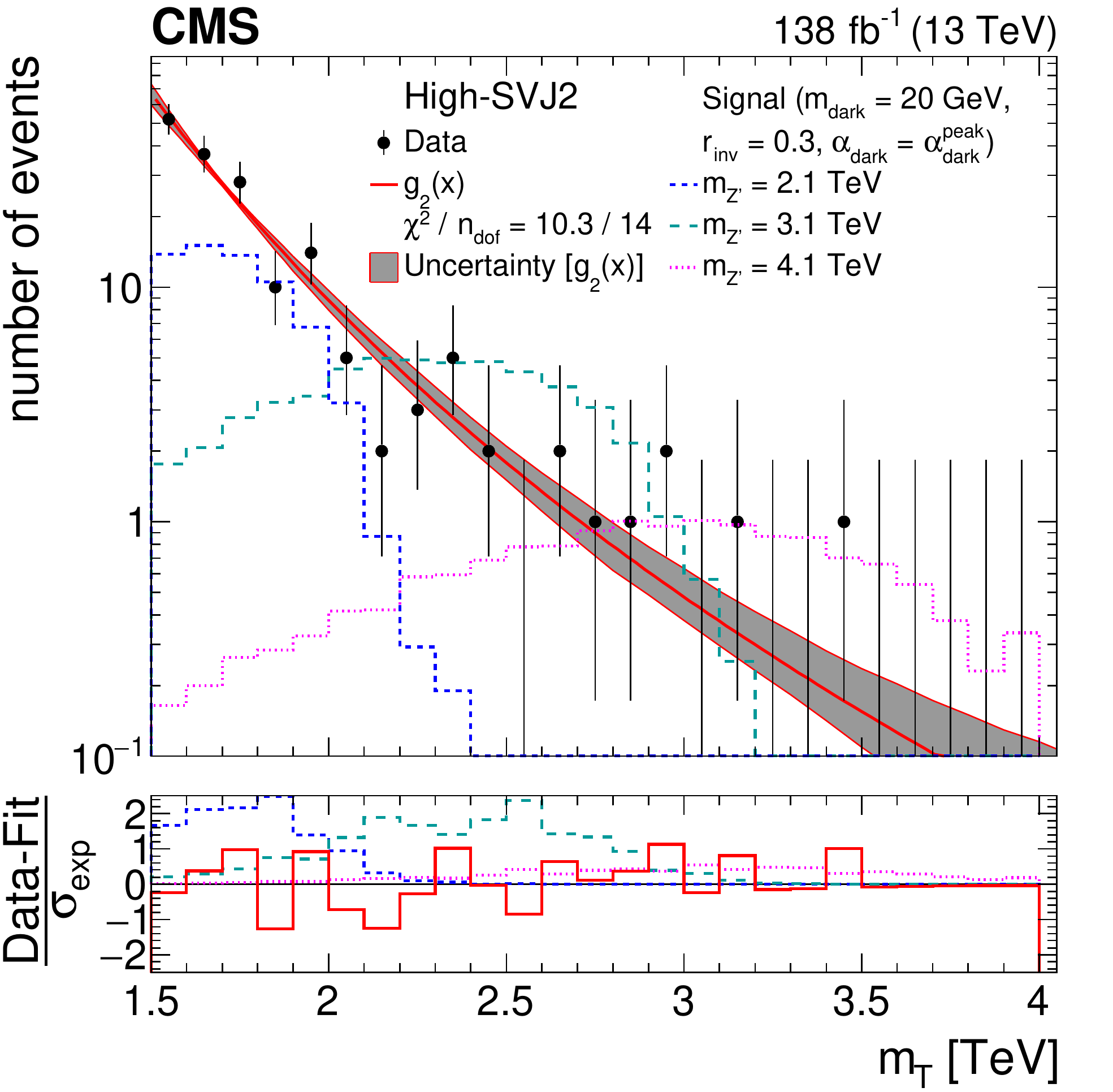}}
\raisebox{0.0\height}{\includegraphics[width=0.37\textwidth]{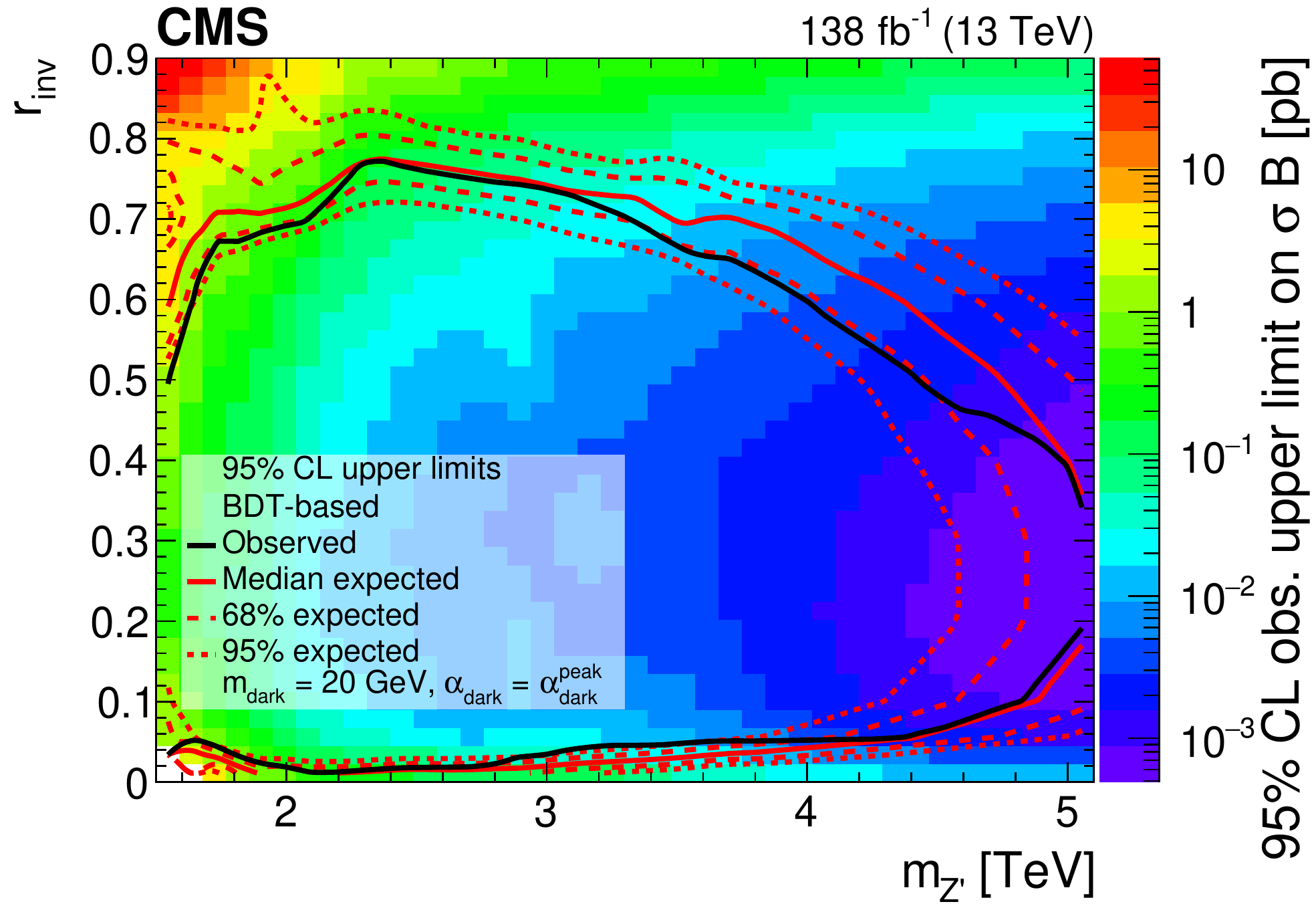}}
\vspace{-0.3cm}
\caption[]{ Distribution of the boosted decision tree output illustrating the separation achieved against the QCD background (left)~\cite{CMS:2021dzg}. Result of the unbinned fit performed on the $m_{T}$ distribution to the data, which compares it to the background prediction and also to superimposed signal scenarios (center)~\cite{CMS:2021dzg}. Exclusion limits on the 2D plane $m_{Z'}$ vs $r_{\text{inv}}$ showing the constraints imposed for some fixed values of the rest of the parameters (right)~\cite{CMS:2021dzg}.}
\label{fig:svj:result}
\end{figure}

The event selection focus on selecting two highly energetic fat jets. The main requirement on the two jets is to have a low value of $|\Delta\phi(\vec{p}^{\text{ jet}}_{T},\vec{p}^{\text{ miss}}_{T})|$, suppressing that way backgrounds like $Z+\text{Jets}$ and $W+\text{Jets}$. The QCD background remains high with this selection, therefore, a dedicated machine learning algorithm is utilized to discriminate the semi-visible jets against QCD-like jets. A boosted decision tree is trained by using fifteen input variables related to jet-substructure quantities. Fig.~\ref{fig:svj:result} (left) illustrates the performance of the algorithm used to separate the two kinds of jets.

Even after applying the above technique, QCD continues to be the dominant background source. The variable used to discriminate between the signal and the residual QCD background is the transverse mass computed from the dijet and the $\vec{p}^{\text{ miss}}_{T}$ systems. The ratio between $m_{T}$ and $\vec{p}^{\text{ miss}}_{T}$ is used to further categorize the analysis into two regions with low and high values of this observable. The background is modeled with a smoothly falling function depending on $m_{T}$, similarly as done in the past in searches for heavy dijet resonances.

An unbinned maximum likelihood fit is performed with the background normalization and the function parameters unconstrained. The results of the fit can be seen in Fig.~\ref{fig:svj:result} (center) for one of the signal categories, evidencing a good agreement between the observed data and the background prediction. Only at masses of $Z'$ around $3.5\text{ TeV}$, a small excess with a local significance of roughly two standard deviations can be seen. Limits are then set on the production cross section, and hence translated to constraints on the 2D plane formed by $m_{Z'}$ and the invisible fraction ($r_{\text{inv}}$), as shown in Fig.~\ref{fig:svj:result} (right). One can clearly see there, how this analysis nicely complements the mono-jet search ($r_{\text{inv}}=1$ limit) and searches for dijet resonances ($r_{\text{inv}}=0$ limit), being able to exclude the intermediate range of the $r_{\text{inv}}$ parameter.

\section{Conclusions}\label{sec:conclusions}

A large variety of interesting DM signatures are currently being covered by ATLAS and CMS. Five of the latest DM searches carried out by both collaborations were presented. None of these analyses found evidence of DM production at the LHC. Nevertheless, the huge amount of data still to come, the analysis improvements envisaged, and the new search proposals planned in the near future confer a strong incentive to continue the hunt for DM at the LHC. 

\section*{References}
\bibliography{bibliography}
%
%
%
%

\end{document}